\documentclass[aoas,nameyear,seceqn,dvips]{arximspdf}
\usepackage{graphics}
\doi{10.1214/09-AOAS306}
\volume{4}
\issue{2}
\pubyear{2010}
\firstpage{943}
\lastpage{961}

\makeatletter
\renewcommand{\citet}[1]{\citeauthor{#1} (\citeyear{#1})}
\renewcommand{\citep}[1]{\citeauthor{#1} (\citeyear{#1})}

\def\sfrac#1#2{#1/#2}
\makeatother

\begin{document}
\begin{frontmatter}

\title{A flexible regression model for count data}
\runtitle{Flexible regression for count data}

\begin{aug}
\author[A]{\fnms{Kimberly F.} \snm{Sellers}\corref{}\ead[label=e1]{kfs7@georgetown.edu}} \and
\author[B]{\fnms{Galit} \snm{Shmueli}\ead[label=e2]{gshmueli@rhsmith.umd.edu}}
\runauthor{K. F. Sellers and G. Shmueli}
\affiliation{Georgetown University and University of Maryland}
\address[A]{Department of Mathematics\\
Georgetown University\\
Washington, DC 20057\\
USA\\ \printead{e1}} %adresu isvedimo komanda gale!
\address[B]{Department of Decision, Operations\\
\quad\& Information Technologies\\
Smith School of Business\\
University of Maryland\\
College Park, Maryland 20742\\
USA\\
\printead{e2}}
\end{aug}

% HISTORY:
\received{\smonth{12} \syear{2008}}
\revised{\smonth{10} \syear{2009}}

% ABSTRACT
%
\begin{abstract}
Poisson regression is a popular tool for modeling count data and is
applied in a
vast array of applications from the social to the physical sciences and beyond.
Real data, however, are often over- or under-dispersed and, thus, not
conducive to Poisson regression.
We propose a regression model based on the Conway--Maxwell-Poisson
(COM-Poisson) distribution to address this problem.
The COM-Poisson regression generalizes the well-known Poisson and
logistic regression
models, and is suitable for fitting count data with a wide range of
dispersion levels.
With a GLM approach that takes advantage of exponential family properties,
we discuss model estimation, inference, diagnostics, and interpretation,
and present a test for determining the need for a COM-Poisson
regression over a standard
Poisson regression. We compare the COM-Poisson to several alternatives
and illustrate
its advantages and usefulness using three data sets with varying dispersion.
\end{abstract}

% KEYWORDS
%
\begin{keyword}
\tolerance=0
\kwd{Conway--Maxwell-Poisson (COM-Poisson) distribution}
\kwd{dispersion}
\kwd{generalized linear models (GLM)}
\kwd{generalized Poisson}.
\end{keyword}

\end{frontmatter}

%s1 ###
\section{Introduction} \label{sec-intro}
Regression models are the most popular tool for modeling the
relationship between a response variable and a set of predictors. In
many applications, the response variable of interest is a count, that
is, takes on nonnegative integer values.
For count data, the most widely used regression model is Poisson
regression, while, for binary data, the logistic (or probit) regression
is most applied. Poisson regression is limiting in its
variance assumption, namely, that for observation $i$ ($i=1, \ldots, n$),
$\operatorname{var}(Y_i) = E(Y_i)$. Even
with the best of intent, however, count data often demonstrate over- or
under-dispersion compared to the Poisson model.

One way to model over-dispersed count data is to use mixture models,
for example, the gamma--Poisson mixture, where Poisson variables have
means $\mu_i$ that follow a gamma distribution. This yields a negative
binomial marginal distribution of the form
\[
P(Y_i=y_i \vert\mu_i,r)= \biggl( \frac{r}{r+\mu_i} \biggr)^r \frac{\Gamma
(r+y_i)}{\Gamma(y_i+1)\Gamma(r)} \biggl(\frac{\mu_i}{r+\mu_i}
\biggr)^{y_i},  \qquad    y_i=0,1,2,\ldots,
\]
where $r \ge0$ and $\mu_i \ge0$ for all $i$ ($i=1,\ldots,n$). The
negative binomial likelihood can be expressed in the form of a
generalized linear model for constant $r$, and a log-link function
($\log\mu_i = \bolds{\beta}' \mathbf{X}_i$) is typically used. Although
negative binomial regression is available in many statistical software
packages, it is limited to modeling only over-dispersed data. In
addition to its inability to fit under-dispersed data, \citet{mccullagh} note that this procedure is ``an unpopular option with a
problematic canonical link.''

An alternative model which can capture both over- and under-dispersion
is the restricted generalized Poisson regression (RGPR) model by \citet{famoye}. The model is given by
\begin{eqnarray}
P(Y_i = y_i \vert\mu_i, \alpha) =  \biggl( \frac{\mu_i}{1 + \alpha\mu_i} \biggr)^{y_i} \frac{(1+ \alpha y_i)^{y_i-1}}{y_i!}
\exp \biggl({\frac{-\mu_i (1+ \alpha y_i)}{1 + \alpha\mu_i}} \biggr),\nonumber
\\
\eqntext{y_i = 0, 1, 2,\ldots,}
\end{eqnarray}
where $\log\mu_i = \bolds{\beta}' \mathbf{X}_i$. It is called a
``restricted'' model, because the dispersion parameter $\alpha$ is
restricted to $1+ \alpha\mu_i > 0$ and $1+\alpha y_i > 0$ [\citep{Cui2006}]. When $\alpha=0$, the model reduces to the Poisson case;
$\alpha>0$ indicates over-dispersion; and $-2/\mu_i<\alpha<0$ indicates
under-dispersion. While this model allows for under- or over-dispersion
in the data (albeit a limited degree of under-dispersion), it belongs
to an exponential family only for a constant dispersion parameter,
$\alpha$. Thus, a more general model with observation-specific
dispersion ($\alpha_i$) will no longer belong to the exponential
family. In short, for count data that are not binary nor follow a
Poisson distribution, readily available, computationally efficient,
flexible regression models are scarce. The need for such a model exists
in many fields where count models are routinely fit to an array of data
sets of varying dispersion.

In this paper we propose using a more general count distribution that
captures a wide range of dispersion. A two-parameter generalized form
of the Poisson distribution, called the Conway--Maxwell-Poisson
(COM-Poisson) distribution [\citep{shmueli}], is
sufficiently flexible to describe a wide range of count data
distributions. It includes as special cases the Poisson, Bernoulli,
and geometric distributions, as well as distributions with dispersion
levels between these three well-known cases (governed by the dispersion
parameter). The COM-Poisson distribution belongs to the exponential family
and therefore possesses advantages in terms of estimation, conjugate
priors, etc. These advantages have proven useful in several
applications, such as using the COM-Poisson sufficient statistics for
purposes of data disclosure [\citep{Kadane2006}], in marketing
applications [\citep{boatwright}, \citep{borle}], and online auctions [\citep{borle2006}]. We describe the COM-Poisson distribution and introduce a
few additional COM-Poisson formulations in Section~\ref{sec-COM}.

In Section~\ref{sec-Reg} we use the COM-Poisson distribution to
formulate a regression model. We discuss model estimation, inference,
interpretation, and diagnostics; obtaining fitted values; and testing
for dispersion. A Bayesian regression formulation using COM-Poisson has
been used in marketing applications by
\citeauthor{borle} (\citeyear{borle,borle2007}), \citet{borle2006}, \citet{boatwright} and \citet{kalyanam}. In each of these
studies $\log(\bolds{\lambda})$ was modeled as a linear function of
predictors, and MCMC was used for estimation. Each of the data sets
included a few thousand observations. For each model, estimation time
was between 2--24 hours. \citet{Lord2008}, motivated by traffic
modeling, used a slightly different Bayesian formulation with $\log
({\bolds{\lambda}}^{1/\nu})$ as the link function. They use
noninformative priors and their model yields good fit. The formulation
used, however, does not take full advantage of the exponential family
features of the COM-Poisson distribution and, in particular, requires
computationally expensive MCMC for estimation. We, instead, approach
the COM-Poisson distribution from a GLM perspective, carefully choosing
a link function (namely, $\log\lambda$) that is advantageous in terms
of estimation, inference, and diagnostics. Our formulation also creates
a generalization of the ordinary Poisson regression as well as logistic
regression, thereby including and bridging two very popular and
well-understood models. Although the logistic regression is a limiting
case ($\nu\rightarrow\infty$), in practice, fitting a COM-Poisson
regression to binary data yields estimates and predictions that are
practically identical to those from a logistic regression.

To show the practical usefulness of the COM-Poisson regression, we
compare its performance to a few alternative regression models:
Poisson, negative binomial, logistic, and RGPR. Section~\ref{sec-examples} considers two data sets of different size and with
different levels of dispersion. Using these data, we illustrate the
advantages of the COM-Poisson model in terms of model fit, inference,
and wide applicability. In Section~\ref{sec-crash} we consider the
\citet{Lord2008} motor vehicle accidents example. We compare the models
along with our COM-Poisson formulation to the Bayesian formulation.
Section~\ref{sec-discuss} concludes with discussion and future directions.

%s2 ###
\section{The COM-Poisson distribution} \label{sec-COM}
The COM-Poisson probability distribution function [\citep{shmueli}] takes
the form
\begin{eqnarray*}
P(Y_i=y_i) = \frac{\lambda_i^{y_i}}{(y_i!)^\nu Z(\lambda_i,\nu)},\qquad
y_i=0,1,2,\ldots,     i = 1, \ldots, n,
\end{eqnarray*}
for\vspace*{-3pt} a random variable $Y_i$, where $Z(\lambda_i, \nu) = \sum
_{s=0}^{\infty}
\frac{\lambda_i^s}{(s!)^\nu}$ and $\nu\ge0$. The ratio between the
probabilities of two
consecutive values is then $\frac{P(Y_i=y_{i}-1)}{P(Y_i=y_i)} = \frac
{y_i^\nu}{\lambda_i}.$ The COM-Poisson distribution generalizes the
Poisson distribution in that the ratio is not necessarily linear in $y_i$,
thereby leading to longer or shorter tails for the distribution. The
COM-Poisson distribution includes three well-known distributions as
special cases: Poisson ($\nu=1$), geometric ($\nu=0, \lambda_i<1$), and
Bernoulli $ ( \nu\rightarrow\infty\mbox{ with probability } \frac
{\lambda_i}{1+\lambda_i}  )$.

In \citet{shmueli} the moments are given in the form
%
%e1 ###
\begin{eqnarray} \label{eq:moments}
E(Y_i^{r+1}) =
\cases{
\lambda_i [E(Y_i+1)]^{1-\nu},                                            &\quad $r=0$,\cr
\displaystyle\lambda_i \frac{\partial}{\partial\lambda_i} E(Y_i^r) +E(Y_i)E(Y_i^r),   &\quad $r>0$,
}
\end{eqnarray}
and the expected value is approximated by
%
%e2 ###
\begin{equation} \label{eq:approxEY}
E(Y_i) = \lambda_i \frac{\partial\log Z(\lambda_i, \nu)}{\partial
\lambda_i} \approx\lambda_i^{1/\nu} -
\frac{\nu- 1}{2\nu}.
\end{equation}
In practice, the expected value can be evaluated by either (1)
estimating the probability density function and truncating the infinite
sum \citep{Minka2003}, or (2)~determining $\hat{\lambda}$, $\hat{\nu}$
and using these estimates to compute the approximation in Equation~(\ref{eq:approxEY}). Another useful result\footnote{We thank Ralph Snyder
for providing this result.} regarding this distribution is that $E(Y^\nu
) = \lambda$. Note that the expected value and variance can also be
written in the form
%
%e4 ###
%e3 ###
\begin{eqnarray}\label{eq:mean}
E(Y_i) &=& \frac{\partial\log Z(\lambda_i, \nu)}{\partial\log\lambda_i},  \\\label{eq:var}
\operatorname{var}(Y_i) &=& \frac{\partial E(Y_i)}{\partial\log\lambda_i} .
\end{eqnarray}
We apply the results from equations (\ref{eq:mean}) and (\ref{eq:var})
to formulate the estimating equations (available in the online
supplemental materials) and the Fisher information matrix (Section~\ref{sec-Reg}).

%s3 ###
\section{Regression formulation} \label{sec-Reg}
Our proposed COM-Poisson regression formulation begins as a
generalization of an ordinary Poisson regression.\break
\citet{mccullagh} view Poisson regression as a special case of loglinear models taking the form
\begin{eqnarray*} \label{model}
\log E(Y_i) = \log\mu_i = \eta_i = \bolds{\beta}' \mathbf{X}_i = \beta_0 +
\beta_1X_{i1} + \cdots+ \beta_pX_{ip},   \qquad  i=1,\ldots,n,
\end{eqnarray*}
where
$\operatorname{var}(Y_i) = \sigma^2 E(Y_i),$ and where $\sigma^2$ denotes the
dispersion parameter [$\sigma^2 > 1$ ($<$1) for over-
(under) dispersion]. Further, they argue that the link function is more
important than the variance
assumption. We will show that, while in some cases dispersion might not
significantly affect mean predictions, it does affect the conditional
distributions and can affect inference.

We can write a similar approximate type of relationship between the
mean and variance via the COM-Poisson distribution.
Using equations (\ref{eq:moments})--(\ref{eq:approxEY}), we can write
(suppressing subscript $i$)
\begin{eqnarray*}
\operatorname{var}(Y) = \lambda\frac{\partial}{\partial\lambda} E(Y) \approx
\lambda\frac{\partial}{\partial\lambda}
\biggl(\lambda^{1/\nu} - \frac{\nu-1}{2\nu}  \biggr) =
\frac{1}{\nu}\lambda^{1/\nu} \approx\frac{1}{\nu} E(Y),
\end{eqnarray*}
in accordance with \citet{mccullagh}. Thus, we can see the relationship
between $\nu$ (or $\frac{1}{\nu} $) and the direction
of data dispersion.

In the following we take a more direct approach to modeling the
dispersion by extending the GLM formulation to the COM-Poisson case and
modeling the relationship between $Y$ and the predictors $\mathbf{X}$ via
a function of $E(Y)$. Although typical link functions are direct
functions of $E(Y)$ [e.g., $E(Y),  \log E(Y), \operatorname{logit}(E(Y))$], the most natural link function for a COM-Poisson
regression is $\eta(E(\mathbf{Y})) = \log{\bolds{\lambda}}$, modeling
the relationship between $E(\mathbf{Y})$ and $\mathbf{X}$ indirectly. This
choice of function is useful for two reasons. First, it coincides with
the link function in two well-known cases: in Poisson regression, it
reduces to $E(\mathbf{Y})={\bolds{\lambda}}$; in logistic regression,
where $\mathbf{p}=\frac{\bolds{\lambda}}{\mathbf{1}+\bolds{\lambda}}$, it reduces to
$\operatorname{logit}(\mathbf{p})=\log\bolds{\lambda}$. The second advantage
of using $\log\bolds{\lambda}$ as the link function is that it
leads to elegant estimation, inference, and diagnostics. This result
highlights the lesser role that the conditional mean plays when
considering count distributions of a wide variety of dispersion levels.
Unlike Poisson or linear regression, where the conditional mean is
central to estimation and interpretation, in the COM-Poisson regression
model, we must take into account the entire conditional distribution.

%s3.1 ###
\subsection{Model estimation} \label{subsec-estimation}
We write the log-likelihood for observation $i$ as
%
%e5 ###
\begin{equation} \label{l_i}
\log L_i(\lambda_i,\nu| y_i) = y_i \log\lambda_i - \nu\log y_i! -
\log Z(\lambda_i,\nu).
\end{equation}
Summing over $n$ observations, the log-likelihood is given by
%
%e6 ###
\begin{equation} \label{eq:logl}
\log L = \sum_{i=1}^n y_i \log\lambda_i - \nu\sum_{i=1}^n \log y_i! -
\sum_{i=1}^n \log Z(\lambda_i,\nu).
\end{equation}

Maximum likelihood coefficient estimates can be obtained by directly
maximizing equation~(\ref{eq:logl}) under the constraint $\nu\geq0$,
using a constrained nonlinear optimization tool (e.g., \texttt{nlminb} in~R). An alternative is to write the log-likelihood as a function
of $\log\nu$, and then maximize it using an ordinary nonlinear
optimization tool (e.g., \texttt{nlm} in~R). A third option for
obtaining the maximum likelihood estimates is to use the GLM framework
to formulate the likelihood maximization as a weighted least squares
procedure (see online supplemental material) and to solve it iteratively.

The GLM formulation is also used for deriving standard errors
associated with the estimated coefficients. The latter are derived
using the Fisher information matrix. For estimating $\bolds{\beta}$ and $\nu$, we have a block Information matrix of the form
%
%e8 ###
%e7 ###
\begin{eqnarray}\label{eq:Fisher}
\mathbf{I} =
\pmatrix{
\mathbf{I}^{\bolds{\beta}}          & \mathbf{I}^{\bolds{\beta},\nu}\vspace*{2pt}\cr
\mathbf{I}^{\bolds{\beta},\nu}      & I^{\nu}
},
\end{eqnarray}
where $\mathbf{I}^{\bolds{\beta}}$ pertains to the estimated
variances and covariances of $\hat{\bolds{\beta}}$, $I^{\nu}$
contains the estimated variance for $\hat{\nu}$, and $\mathbf{I}^{\bolds{\beta},\nu}$ contains the componentwise estimates of the
covariance between $\hat{\bolds{\beta}}$ and $\hat{\nu}$. Details
regarding the information matrix components are available in the online
supplementary material. R code for estimating COM-Poisson
regression coefficients and standard errors is available at
\href{http://www9.georgetown.edu/faculty/kfs7/research}{www9.georgetown.edu/faculty/kfs7/research}.

%s3.2 ###
\subsection{Testing for dispersion} \label{sec-testing}
How much data dispersion should exist to warrant deviation from Poisson
regression? The set of hypotheses, $H_0\dvtx  \nu= 1$ vs.~$H_1\ :  \nu\ne1$, ask whether the use of Poisson regression is reasonable versus
the alternative of fitting COM-Poisson regression. Note that $H_1$ does
not specify the direction (over vs. under) of data dispersion. This can
be assessed, however, via exploratory data analysis and the dispersion
estimate, $\hat{\nu}$, from the fitted COM-Poisson regression.

We derive the test statistic,
\[
C = -2 \log\Lambda= -2  \bigl[\log L  \bigl(\hat{\bolds{\beta}}{}^{(0)}, \hat{\nu}=1  \bigr) - \log L  (\hat{\bolds{\beta}},\hat{\nu}  ) \bigr],
\]
where $\Lambda$ is the likelihood ratio test statistic, ${\hat{\bolds{\beta}}}{}^{(0)}$ are the maximum likelihood estimates
obtained under $H_0\dvtx  \nu=1$ (i.e., the Poisson estimates), and $(\hat
{\bolds{\beta}},\hat{\nu})$ are the maximum likelihood estimates
under the general state space for the COM-Poisson distribution. Under
the null hypothesis, $C$ has an approximate $\chi^2$ distribution with
1 degree of freedom. For small samples, the test statistic distribution
can be estimated via bootstrap.

%s3.3 ###
\subsection{Computing fitted values}
\label{subsec-fitted}
Once a COM-Poisson regression model has been estimated, we can obtain
fitted values ($\hat{y}_i$) in one of two ways:
\begin{enumerate}
\item Estimated means: We can use the approximation in equation~(\ref{eq:approxEY}) and obtain fitted values by
$\hat{y}_i | \mathbf{x}_i = \hat{\lambda}_i^{1/\hat{\nu}} - \frac{\hat{\nu
}-1}{2\hat{\nu}}$,
where $\hat{\lambda}_i = \exp(\mathbf{x}_i'\hat{\bolds{\beta}})$.
Note that this approximation is accurate for $\nu\leq1$ or $\lambda_i>10^{\nu}$ [\citep{Minka2003}].

\item Estimated medians: When the mean approximation is inadequate (or
in general), we can obtain percentiles of the fitted distribution by
using the inverse-CDF for $\hat{y}_i|\mathbf{x}_i$ and $\hat{\nu}$. In
particular, we use the estimated median to obtain fitted values.
\end{enumerate}

%s3.4 ###
\subsection{Model inference} \label{sec-inference}
Due to the GLM formulation, the statistical significance of individual
predictors can be obtained by using the asymptotic standard normal
distribution of $\sfrac{\hat{\beta}_j}{\hat{\sigma}_{\hat{\beta}_j}}$.
In the case of small samples, however, where the asymptotic normality
might not hold (as in other count data regression models),
bootstrapping can be used to estimate the distributions of the
coefficients of interest. With small samples, COM-Poisson model
estimation is very fast, thereby being practically useful for bootstrap.

A parametric COM-Poisson bootstrap can be implemented by resampling
from a COM-Poisson distribution with parameters $\hat{\bolds{\lambda}}=\exp(\mathbf{X}' \hat{\bolds{\beta}})$ and $\hat{\nu}$, where $\hat
{\bolds{\beta}}$, $\hat{\nu}$ are estimated from a COM-Poisson
regression on the full data set. The resampled data sets include new
$Y$ values accordingly. Then, for each resampled data set, a
COM-Poisson regression is fit, thus producing new associated estimates,
which can then be used for inference.

%s3.5 ###
\subsection{Coefficient interpretation}
There are two main approaches for interpreting coefficients in
regression models [\citep{Long1997}]. One examines changes in the
conditional mean for a unit increase in a single predictor, for
example, $E(Y|X_j=x_j, \mathbf{X}_{i\neq j}=\mathbf{x})$ and
$E(Y|X_j=x_j+1, \mathbf{X}_{i\neq j}=\mathbf{x})$. In additive
models, such as a linear regression, the difference between the two
conditional means [or the derivative of $E(Y|X)$ with respect to $X_j$]
is used for interpretation [``a unit increase in $X_j$ is associated
with a $\beta_j$ increase in $E(Y)$'']; in multiplicative models, such
as the Poisson or logistic regressions, the ratio of the two
conditional means is used for interpretation [``a unit increase in
$X_j$ is associated with a factor of $e^{\beta_j}$ increase in $E(Y)$
or the odds'']. The second approach, which is used for coefficient
interpretation in other types of nonlinear regression models (e.g.,
probit regression), is to directly examine the relationship between
fitted values and changes in a predictor. This can be done via
graphical plots for less than two predictors, while, for more than two
predictors, there are various solutions such as fitted value
consideration at selected values of the predictors.

In the COM-Poisson regression case, we cannot use the first approach
that compares conditional means directly, because the relationship
between the conditional mean and the predictors is neither additive nor
multiplicative (except for the special cases of Poisson and logistic
regressions). For example (considering a single predictor model), the
ratio of conditional means leads to a complicated nonlinear
relationship between a unit increase in $X$ and the effect on $E(Y|X)$.
However, the result $E(Y^\nu) = \lambda$ in Section~\ref{sec-COM}
indicates a multiplicative relationship between the predictors and
$E(Y^\nu)$. It appears, however, that interpreting the effect of
individual predictors on the conditional mean (or median) directly is
most straightforward via the second approach.

Because coefficients from a COM-Poisson regression model are on a
different scale than those from an ordinary Poisson model, for purposes
of crude comparison, one can simply divide the COM-Poisson coefficients
by $\nu$. This approach is reasonable because $E({\mathbf{Y}}^\nu)={\bolds{\lambda}}$.

%s3.6 ###
\subsection{Model diagnostics}
Due to the GLM formulation and, in particular, the IWLS framing (see
online supplemental material), standard GLM diagnostics can be used for
residual analysis of a fitted COM-Poisson regression model. We use the
matrices $\mathcal{W}$ and $\mathcal{X}$ as defined there for computing
leverage, and the popular Pearson and Deviance residuals.
Leverage can be computed from the hat matrix, $H = \mathcal{W}^{1/2} \mathcal{X}  (\mathcal{X' W X} )^{-1} \mathcal{X}' \mathcal{W}^{1/2}$. An
observation with an unusually high value of $h_i$ is suspect of having
influence (although $H$, like other nonlinear models, depends on the
estimated parameters). Meanwhile, using ordinary GLM formulations, we
can write the Pearson residual for observation $i$ [\citep{Davison1992}]
as $r_{P,i} = \frac{Y_i-\hat{\mu_i}}{\sqrt{w_i(1-h_i)}}$, where $\hat{\mu}_i = \widehat{E(Y_i)}$, and the standardized deviance residual for
observation $i$ can be written as $r_{D,i} = \operatorname{sgn}(Y_i-\hat{\mu
}_i)\frac{d_i}{\sqrt{1-h_i}}$, where $d_i = -2[\log L(\hat{\mu}_i, y_i;
\hat{\nu}) - \log L(y_i, y_i; \hat{\nu})].$ These two types of
residuals can be computed directly or approximated using the mean
approximation in Equation~(\ref{eq:approxEY}). In particular, for
deviance residuals, the approximation leads to
%
%e9 ###
\begin{eqnarray}\label{eq:DevApprox}
d_i &=& 2  \biggl[ y_i \hat{\nu} \log
\biggl( {\biggl(y_i + \frac{\hat{\nu }-1}{2\hat{\nu}}\biggr)}\Big/
{\biggl(\hat{\mu}_i + \frac{\hat{\nu}-1}{2\hat{\nu}} \biggr)}\biggr)\nonumber
\\[-8pt]\\[-8pt]
&&\phantom{2  \biggl[}{}+\log \biggl( { Z \biggl(  \biggl(\hat{\mu}_i + \frac{\hat{\nu}-1}{2\hat{\nu}} \biggr)^{\hat{\nu}}, \hat{\nu} \biggr) }\Big/
{ Z \biggl(\biggl(y_i + \frac{\hat{\nu}-1}{2\hat{\nu}} \biggr)^{\hat{\nu}}, \hat{\nu} \biggr) }  \biggr)  \biggr].\nonumber
\end{eqnarray}
The existence of equation (\ref{eq:DevApprox}) is constrained in that $Y >
k$ for $\hat{\nu} < \frac{1}{2k +1}$; $k \in\mathbf{N}^+$. We can,
however, modify equation (\ref{eq:DevApprox}) in order to obtain valid
results for $d_i$. For example, when $\nu< 1$ and $Y=0$, we set $Z
( (y_i + \frac{\hat{\nu}-1}{2\hat{\nu}} )^{\hat{\nu}}, \hat{\nu
} ) = 1$. Another option is to use the exact deviance equations
supplied above, though this is computationally more expensive. Finally,
while the approximation is accurate for $\lambda>10^\nu$ or $\nu<1$, we
have found that deviance residuals computed using equation (\ref{eq:DevApprox}) are quite accurate even outside that range (e.g., for
under-dispersed data with low counts).

A probability plot of the deviance residuals, as well as a scatter plot
of $\log(\hat{\lambda})$ versus deviance residuals, can help assess
model adequacy and detect outliers. Although normal probability plots
are common, deviance residuals for nonlinear models can be far from
normally distributed [\citep{Ben2004}]. One alternative is to ignore the
fit to normality on the normal probability plot, and use it just to
detect outliers. Another option is to use bootstrap to estimate the
distribution of deviance residuals, and then to create a QQ plot of the
deviance residuals against their estimated distribution.

%s4 ###
\section{Examples}\label{sec-examples}
In this section we fit regression models to data sets characterized by
under-dispersion, and with binary outcomes (i.e., extreme
under-dispersion); Section~\ref{sec-crash} discusses the
over-dispersion example considered by \citet{Lord2008}. We fit various
popular regression model choices for count data: Poisson, negative
binomial (NB), restricted generalized Poisson (RGPR), and COM-Poisson.
For the binary data set, we also fit a logistic regression. The goal of
this section is to compare the COM-Poisson to the other models in terms
of fit, inference, and flexibility. The small sample size and dimension
of the first data set is useful for directly observing the effect of
dispersion. In particular, we show the effect of dispersion on the
conditional distribution of fit. We evaluate goodness-of-fit and
predictive power by examining the fitted values and comparing values of
MSE and AIC$_C$ (the Akaike Information Criterion\footnote{All models
aside from Poisson have a penalty term in the AIC$_C$ that takes into
account the extra dispersion parameter.} corrected for small sample
size) across models.

Note that, except for the Poisson and logistic regressions, the other
models considered have an extra dispersion parameter that is assumed
fixed across observations, but unknown. Each of the models is estimated
by maximum likelihood. The Poisson, NB, and logistic regressions are
estimated using ordinary GLM functions in R. COM-Poisson is
estimated using nonlinear optimization in R, and standard errors
are estimated as described in Section~\ref{subsec-estimation}. RGPR is
estimated using constrained nonlinear optimization in R and
standard errors are estimated as described in \citet{famoye}.

%s4.1 ###
\subsection{Regression with under-dispersed data: Airfreight breakage}
We first consider the airfreight breakage example from [\citet{Kutner}, page~35, Exercise~1.21] where data are given on 10 air shipments,
each carrying 1000 ampules on the flight. For each shipment $i$, we
have the number of times the carton was transferred from one aircraft
to another ($X_i$) and the number of ampules found broken upon arrival
($Y_i$). The data are provided online among the supplementary material.

%t1 ###
\begin{table}[b]
\caption{Estimated coefficients and standard errors (in parentheses)
for the airfreight example, for various regression models. NB and
Poisson regression produce the same estimates. The RGPR did not converge}\label{tab-Betas}
\begin{tabular*}{\textwidth}{@{\extracolsep{\fill}}lcc@{}} \hline
\textbf{Model} & $\bolds{\hat{\beta}_0   (\hat{\sigma}_{\hat{\beta}_0})}$ & $\bolds{\hat{\beta}_1   (\hat{\sigma}_{\hat{\beta}_1})}$ \\
\hline
Poisson/NB & \phantom{0}2.3529 (0.1317) & 0.2638 (0.0792) \\
COM-Poisson ($\hat{\nu}=5.7818, \hat{\sigma}_{\hat{\nu}}=2.597$) &
13.8247 (6.2369)& 1.4838 (0.6888) \\
\hline
\end{tabular*}
\end{table}

We first estimated the COM-Poisson regression coefficients and tested
for dispersion. The estimated dispersion parameter is $\hat{\nu}=5.78$,
indicating under-dispersion. To test for dispersion, we use parametric
bootstrap (see Section~\ref{sec-inference}) rather than the dispersion
test, due to the small sample size. The 90\% bootstrap confidence
interval for $\nu$ is (4.00, 21.85), indicating dispersion that
requires a COM-Poisson regression instead of ordinary Poisson
regression. We proceed by attempting to fit the four regression models.
The estimated coefficients and standard errors for three of these
models (Poisson, NB, and COM-Poisson) are given in Table~\ref{tab-Betas}; NB regression produces identical estimates to that from
Poisson regression. RGPR did not converge and, therefore, no estimated
model is produced. This highlights the limited ability of RGPR to fit
under-dispersed data. In general, for under-dispersed data, the RGPR
probability function ``gets truncated and does not necessarily sum to
one'' [\citep{Famoye2004}]. This example appears to fall exactly under
this limitation.

\begin{figure}

\includegraphics{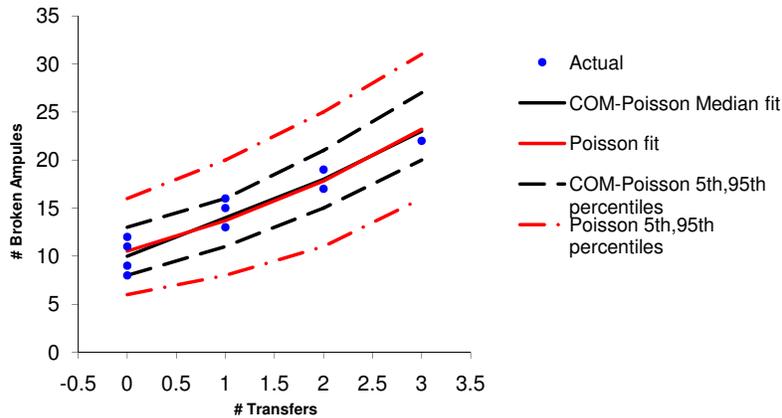}

\caption{Fitted mean curves (solid lines), 5th and 95th percentile
curves (broken lines) for Poisson and COM-Poisson regression models for
the airfreight breakage data (dots).} \label{fig:AirfreightFit}
\end{figure}

\begin{figure}

\includegraphics{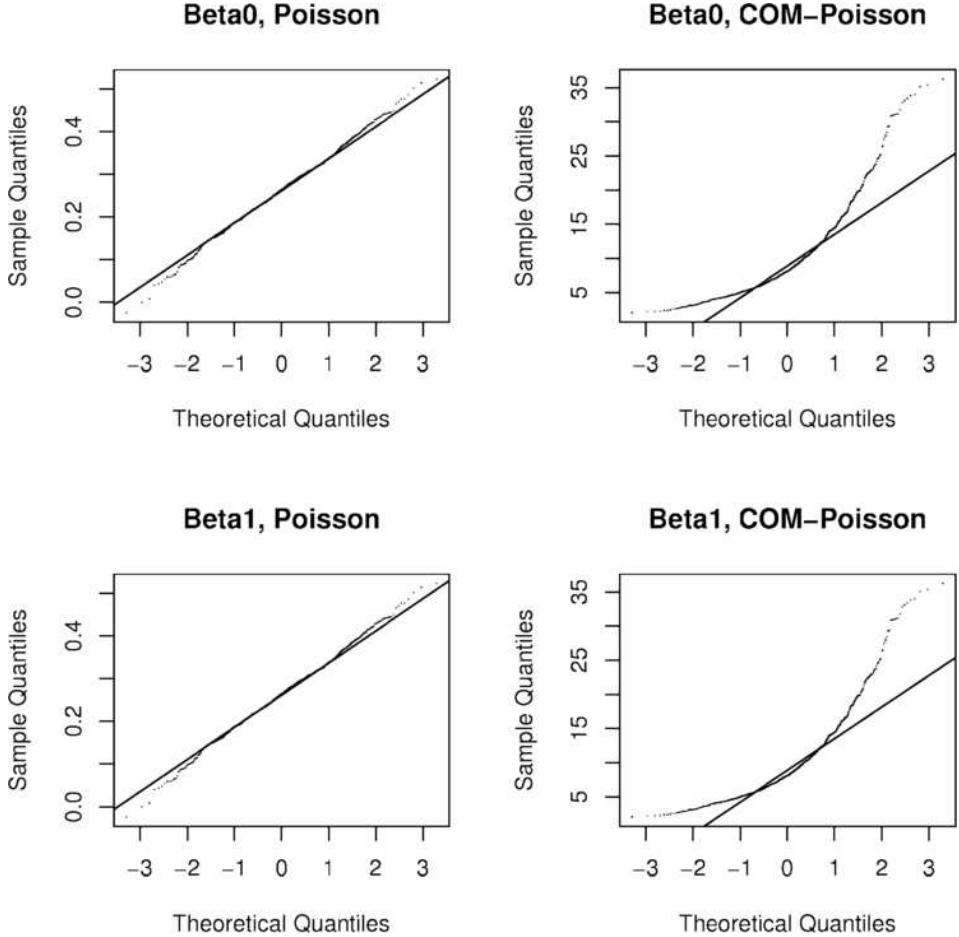}

\caption{Normal probability plots of $\hat{\beta}_0$ (top) and $\hat
{\beta}_1$ (bottom) based on 1000 bootstrap samples of the airfreight
breakage data. Negative binomial estimation produces identical results
to those from Poisson regression. RGPR estimation procedure does not
converge.} \label{fig:FreightBootstrp}
\end{figure}

Fitted values from the models are provided online in the supplementary
material where, for the COM-Poisson, we use the estimated conditional
median for fitted values because the approximation (\ref{eq:approxEY})
is likely to be inaccurate (here, $\nu>1$ and $\lambda\ngtr 10^\nu
$). We find that the models are similar in terms of the fitted values
that they generate (see also Figure~\ref{fig:AirfreightFit}). In terms
of MSE and AIC$_C$, the COM-Poisson shows best fit, although the
differences between models for these values are not large (see Table~\ref{tab-Freightstats}). The similarity of the regression models is
also in terms of the coefficient magnitudes (after dividing the
COM-Poisson coefficients by $\hat{\nu}$). The models differ, however,
in two important ways. First, although the fitted values are similar,
the conditional distribution differs markedly across the models, as can
be seen by comparing the 5th and 95th percentile curves in Figure~\ref{fig:AirfreightFit}. Second, the models initially appear to differ in
terms of inference. Comparing the Poisson, and COM-Poisson estimated
models, we find that the ratio\vspace*{1pt} $\sfrac{\hat{\beta}_1}{\hat{\sigma}_{\hat{\beta}_1}}$ is 3.33 and 2.15, respectively. Due to the small sample
size, however, the normal approximation might not be adequate. We
therefore examined the distributions of $\hat{\beta}_0$ and $\hat{\beta
}_1$ for each of the models, based on 1000 parametric bootstrapped
samples (see Section~\ref{sec-inference}). Figure~\ref{fig:FreightBootstrp} displays normal probability plots for the
estimated coefficients. We see that the distributions for the
COM-Poisson model are skewed. To evaluate statistical significance of
the predictor (number of transfers), we examine the percent of the
distribution of $\hat{\beta}_1$ to the left of the value $\beta_1=0$.
In both models, this percent is zero, indicating high statistical significance.

%f1 ###

%t2 ###
\begin{table}[b]
\tablewidth=6cm
\caption{Airfreight breakage example: goodness-of-fit and predictive power statistics} \label{tab-Freightstats}
\begin{tabular*}{\tablewidth}{@{\extracolsep{\fill}}lcc@{}}
\hline
& \textbf{COM-Poisson} & \textbf{Poisson} \\
& \textbf{median fit} & \textbf{fit} \\
\hline
AIC$_C$ & \textbf{47.29}\phantom{0} & 52.11 \\
MSE & \phantom{0}\textbf{1.90} & \phantom{0}2.21 \\
\hline
\end{tabular*}
\end{table}

In terms of model interpretation, the Poisson regression indicates that
a unit increase in the number of transfers is associated with a factor
increase of 1.3 in the average number of broken ampules. Looking at
Figure~\ref{fig:AirfreightFit}, however, shows that interpretations in
terms of the average number of broken ampules is insufficient. In
particular, the number of transfers seems to affect the entire
distribution of the number of broken ampules, as indicated by the
fitted COM-Poisson model. Indeed, the COM-Poisson curves in Figure~\ref{fig:AirfreightFit} can be used directly for interpreting the
relationship between number of transfers and number of broken ampules.

%f2 ###

\begin{figure}

\includegraphics{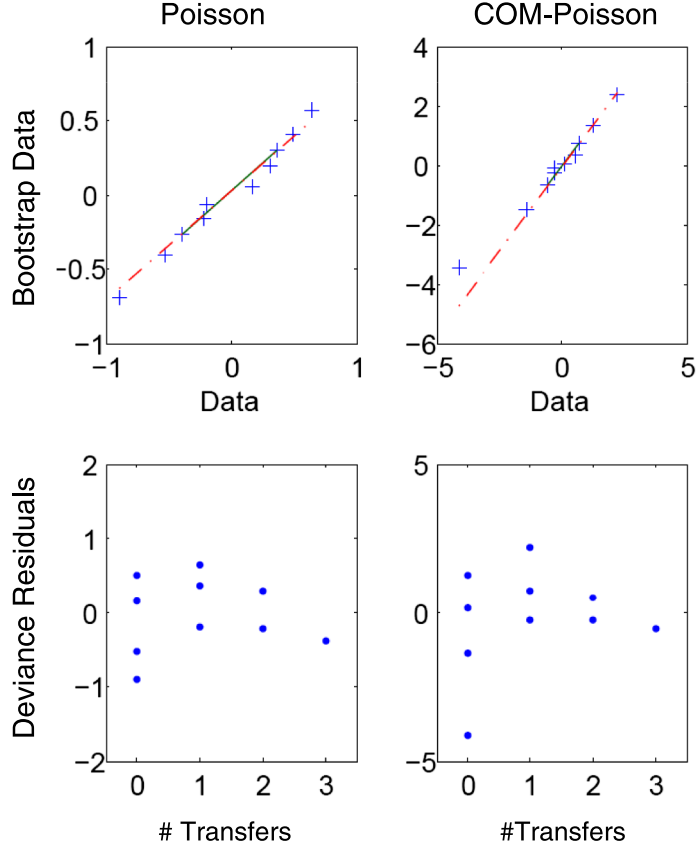}

\caption{QQ plots of the scaled deviance residuals (top) and
scatterplots of the scaled deviance residuals vs. the predictor
(bottom) for the airfreight breakage data. Each column corresponds to a
different regression model.} \label{fig:DevResFreight}
\end{figure}

Finally, we examine leverage and scaled deviance residuals from each of
the models. Figure~\ref{fig:DevResFreight} displays scatterplots of the
deviance residuals versus the single predictor (which is equivalent to
plotting versus $\log\hat{\lambda}$ for the Poisson and COM-Poisson
models), and QQ plots. Leverage values are available in the online
supplementary materials. Overall, there is no noticeable pattern in any
of the scatterplots. Both models indicate observation \#5 (with $X=3$)
as suspect of being influential, and observation \#7 as an outlier
(having a large negative deviance residual), particularly for the
COM-Poisson model.

%f3 ###

%s4.2 ###
\subsection{Regression with extreme under-dispersion: Book purchases}
We now consider the case where the outcome variable is binary, and
where typically a logistic regression would have been considered.
Although the logistic regression is theoretically only a limiting case
of the COM-Poisson regression, we show that (in practice) a fitted
COM-Poisson to binary outcome data produces practically identical
results to a logistic regression. We use a data set from \citet{Lattin2003} that describes the results of a direct-marketing campaign
by a book club, for a certain art book.\footnote{Two additional examples
where COM-Poisson regression is applied to binary data (showing similar
results) are given in the online supplemental materials.} The data set
contains the results for 1000 customers. The outcome is whether the
customer purchased the art book or not. The two predictor variables are
the number of months since the customer's last purchase (\textit{Months}), and
the number of art books that the customer has purchased in the past
(\textit{ArtBooks}). We use this data set to show the flexibility of the
COM-Poisson regression over the alternatives discussed above. In
particular, we show that the COM-Poisson regression produces estimates
and predictions that are identical (to multiple decimals) to those from
a logistic regression, and that RGPR and NB fail to converge altogether.

Table~\ref{tab-Books} provides the parameter estimates from the
Poisson, logistic, and COM-Poisson regression models, respectively. The
NB regression estimates are identical to the Poisson estimates. RGPR is
absent from Table~\ref{tab-Books} because it has limited ability to
capture under-dispersion, thus, it fails to converge.

%t3 ###
\begin{table}
\caption{Estimated coefficients and standard errors (in parentheses)
for Book Club example, for four regression models (NB estimates are
identical to Poisson; RGPR did not converge). The estimates for the
logistic and COM-Poisson models are identical, even to eight decimal
places}\label{tab-Books}
\begin{tabular*}{\textwidth}{@{\extracolsep{\fill}}lccc@{}}
\hline
\textbf{Model}
& $\bolds{\hat{\beta}_0   (\hat{\sigma}_{\hat{\beta}_0})}$
& $\bolds{\hat{\beta}_{\mathit{Months}}   (\hat{\sigma}_{\hat{\beta}_{\mathit{Months}}})}$
& $\bolds{\hat{\beta}_{\mathit{ArtBooks}}   (\hat{\sigma}_{\hat{\beta}_{\mathit{ArtBooks}}})}$ \\
\hline
Poisson/NB                  & $-$2.29 (0.18)& $-$0.06 (0.02) & 0.73 (0.05)\\
Logistic                    & $-$2.23 (0.24)& $-$0.07 (0.02) & 0.99 (0.14)\\
COM-Poisson                 & $-$2.23 (0.24)& $-$0.07 (0.02) & 0.99 (0.14)\\
\quad($\hat{\nu}=30.4, \hat{\sigma}_{\hat{\nu}}=10{,}123$) & & & \\
\hline
\end{tabular*}
\end{table}

With respect to comparing COM-Poisson with logistic regression, it is
clear that the two models produce identical results in terms of
coefficients and standard errors (even to eight decimals). Meanwhile,
we note the large estimated value for $\nu$, along with its broad
standard error. This is in congruence with the terms of the COM-Poisson
distribution for the special case of a Bernoulli random variable
(namely, $\nu\rightarrow\infty$). Furthermore, comparing fitted
values (or predictions), using the estimated COM-Poisson median as the
fitted value (in accordance with Section~\ref{subsec-fitted}) yields
values that are identical to those from a logistic regression with
cutoff value 0.5. To obtain fits for other cutoff values, the
corresponding percentile should be used. Finally, although the Poisson
model does converge, it is clearly inappropriate in terms of inference,
and produces fitted values that are not binary.

%s5 ###
\section{Regression with over-dispersed data: Modeling motor vehicle
crashes} \label{sec-crash}
The previous section shows the flexibility of the COM-Poisson
regression to capture under-dispersion, which exceeds the ability of
models such as the negative binomial and RGPR. We now examine an
over-dispersed data set used by \citet{Lord2008} which contains motor
vehicle crash data in 1995, at 868 signalized intersections located in
Toronto, Ontario. For each intersection, measurements included the
annual number of crashes at the intersection ($Y$) and two traffic flow
variables. See \citet{Lord2008} for further details on the data.

Because motor vehicle crash data contain counts, Poisson and negative
binomial regressions are common models in the field of transportation
safety. For the Toronto data set, \citet{Lord2008} proposed using a
Bayesian COM-Poisson regression formulation to model the
over-dispersion. In particular, they used noninformative priors and
modeled the effect of the two traffic variables on the number of
crashes via the link function $\log ( {\bolds{\lambda}}^{1/\nu
}  ) = {\mathbf{X} \bolds{\beta}}$. Parameter estimation was then
performed via MCMC. The authors note that estimation for this data set
used 35,000 replications, requiring nearly five hours of computation.
Comparing goodness-of-fit and out-of-sample prediction measures, \citet{Lord2008} showed the similarity in performance of the COM-Poisson and
negative binomial regression. They then motivate the advantage of the
COM-Poisson over the negative binomial regression in the ability to fit
under-dispersion and low counts.

The goal of this section is two-fold: (1) to extend the model
comparison in \citet{Lord2008} beyond the negative binomial model to
additional models, as well as to examine a wider range of model
comparison aspects, and (2)~to compare the Bayesian COM-Poisson
formulation to our formulation and show the advantages gained by using
our formulation. Although goodness-of-fit measures might indicate
similarity of the COM-Poisson performance to other models, model
diagnostics provide additional information.

%s5.1 ###
\subsection{Model estimation}
Various regression models were fit to the Toronto intersection crash
data. Following \citet{Lord2008}, the response was the number of
crashes at the intersection, and the two covariates were the two
log-transformed traffic flow variables.

Table~\ref{tab-Crash} displays the estimated models: two COM-Poisson
formulations [our model and the Bayesian model of \citet{Lord2008}],
and three alternative regression models (Poisson, NB, and RGPR). From
$\hat{\nu}<1$ and $\hat{\alpha}>0$, over-dispersion is indicated. All
$\hat{\beta}$ coefficients appear similar across the models. For
standard errors, the Poisson estimates are much smaller than in other
models (as expected in over-dispersion).

%t4 ###
\begin{table}
\tabcolsep=0pt
\caption{Estimated models: comparing two COM-Poisson formulations [ours
and \protect\citet{Lord2008}], and alternative models for the Toronto crash
data. For ease of comparison, we report the COM-Poisson estimates and
standard errors from our formulation in terms of $\hat{\nu}$
multipliers, to reflect the comparable scale with estimates from the
other models} \label{tab-Crash}
{\fontsize{8.7}{10.7}\selectfont
\begin{tabular*}{\textwidth}{@{\extracolsep{\fill}}lllll@{}}
\hline
\textbf{Model}
& \multicolumn{1}{c}{\textbf{Extra parameter}}
& \multicolumn{1}{c}{$\bolds{\hat{\beta}_0   (\hat{\sigma}_{\hat{\beta}_0})}$}
& \multicolumn{1}{c}{$\bolds{\hat{\beta}_1   (\hat{\sigma}_{\hat{\beta}_1})}$}
& \multicolumn{1}{c@{}}{$\bolds{\hat{\beta}_2   (\hat{\sigma}_{\hat{\beta}_2})}$} \\
\hline
Our formulation & $\hat{\nu}=0.3492$ (0.0208) & $-$11.7027$\hat{\nu}$
                          (0.7501$\hat{\nu}$) & 0.6559$\hat{\nu}$ (0.0619$\hat{\nu}$) &
                          0.7911$\hat{\nu}$ (0.0461$\hat{\nu}$)
\\
Lord, Guikema               & $\hat{\nu}=0.3408$ (0.0208) & $-$11.53 (0.4159) &0.6350 (0.0474) & 0.7950 (0.0310) \\
\quad and Geedipally \\
\quad (\citeyear{Lord2008})\\
Poisson                        &                             & $-$10.2342 (0.2838) & 0.6029 (0.0288) & 0.7038 (0.0140) \\
Neg-Bin                        &$\hat{r}=7.154$ (0.625)      & $-$10.2458 (0.4626) & 0.6207 (0.0456)& 0.6853 (0.0215) \\
RGPR                           &$\hat{\alpha}=0.050$ (0.004) & $-$10.2357 (0.4640) & 0.6205 (0.0451)& 0.6843 (0.0215) \\
\hline
\end{tabular*}
}
\end{table}

Comparing the two COM-Poisson formulations, the two are nearly
identical in terms of $\hat{\nu}$ and its standard error [or the
equivalent posterior credible standard error for \citet{Lord2008}], and
in terms of the $\hat{\beta}$ coefficients (after scaling by a factor
of $\hat{\nu}$, due the different formulation of the relationship
between the covariates and the response). These similarities between
the Bayesian and classic formulations indicate that the prior
information does not affect the model, here most likely due to the
large size of the data set. The most dramatic difference between the
two implementations is in run time: our estimation took less than three
minutes, compared to five hours required by the Bayesian MCMC. This
difference has significance especially since \citet{Lord2008} used
noninformative priors to obtain their estimates. Thus, in the absence
of strong prior information or in the presence of a large data set, our
formulation provides more efficient estimation. Even in the presence of
prior information, our method is still useful for obtaining initial
estimates to speed up the MCMC process.

%s5.2 ###
\subsection{Model performance}
Comparing goodness-of-fit measures, the two\break COM-Poisson formulations
are practically identical in terms of $\hat{\beta}$ and thus produce
nearly identical fitted values. Compared to the other regression
models, the COM-Poisson model has lower MSE and AIC values, indicating
better fit and predictive power (see Table~\ref{tab-CrashFit}). The
COM-Poisson dispersion test (with $C=518$, and associated $p$-value${}={}$0)
indicates that the COM-Poisson model is more adequate than Poisson regression.

%t5 ###
\begin{table}[b]
\caption{Goodness-of-fit comparison of COM-Poisson with alternative
fitted models}\label{tab-CrashFit}
\begin{tabular*}{\textwidth}{@{\extracolsep{\fill}}lcccc@{}}
\hline
& \textbf{COM-Poisson}& \textbf{Poisson} & \textbf{Neg-Bin} & \textbf{RGPR}\\
\hline
AIC & \textbf{5073} & 5589 & 5077 & 5092 \\
MSE & \textbf{32.57} & 32.60 & 32.70 & 32.71\\
\hline
\end{tabular*}
\end{table}

%f4 ###
\begin{figure}

\includegraphics{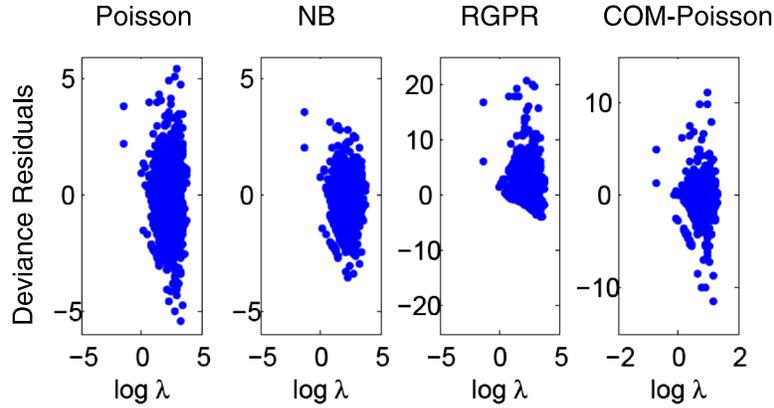}

\caption{Scatterplots of the scaled deviance
residuals vs. $\log\hat{\lambda}$. Each column corresponds
to a different regression model.
For RGPR the deviance residuals are unscaled.}
\label{fig:DevResCrashScatter}
\end{figure}

We now examine model diagnostics to better understand model fit. Figure~\ref{fig:DevResCrashScatter} displays scatterplots of the scaled
deviance residuals vs. $\log\hat{\lambda}$. For RGPR, we use unscaled
deviance residuals (as $H$ is unavailable). From the residual plots and
the leverage measures (available in the online supplementary
materials), we find that the NB model marks nearly half of the $Y=0$
observations as influential, and flags mostly high-count observations.
The Poisson and NB models mark the observations with largest $X$ values
as influential. In contrast, COM-Poisson diagnostics point out eight
observations with large residuals (\#15, \#42, \#247, \#424, \#494, \#618, \#619, \#757) and three with high leverage (\#133, \#801, \#835).
Three of the large-residual intersections have a large number of
crashes with relatively little traffic (small values of the
covariates). The remaining large-residual intersections have a small to
medium number of crashes, but less substantial traffic on one of the
traffic flow covariates. All of these observations are also flagged by
at least one other regression method, with observations \#15 and \#618
being flagged by all methods.

%s5.3 ###
\subsection{Inference}
In terms of drawing inference about the effect of the traffic flow
covariates on the number of crashes, we examine the coefficients and
standard errors and assume a normal approximation. In this case, the
effects are very strong across all models, resulting in $p$-values of
zero for each of the two covariate coefficients.

%s6 ###
\section{Discussion} \label{sec-discuss}
The COM-Poisson regression model provides a practical tool for modeling
count data that have various levels of dispersion. It generalizes the
widely-used Poisson regression, as well as allows for other levels of
dispersion. Using a GLM approach and taking advantage of the
exponential family properties of the COM-Poisson distribution, we
provide a straightforward, elegant, computationally efficient framework
for model estimation, dispersion testing, inference, and diagnostics.
The data examples illustrate the differences and similarities that
arise in practice when using a COM-Poisson regression versus more
traditional regression models. For moderate to high counts, fitted
values can be similar across models but the conditional fitted
distribution can differ markedly. Models also tend to diverge in terms
of inference for single predictors, implying that inappropriate use of
a Poisson model (instead of a COM-Poisson model) can lead to erroneous
conclusions.

One important insight from the COM-Poisson regression model is that, in
a model that allows for different levels of dispersion, the role of the
conditional mean is no longer central. Unlike linear regression or
Poisson regression where the conditional mean is central to
interpretation, the COM-Poisson regression uses a more general function
of the response distribution. The resulting model means that, when
examining goodness-of-fit or when predicting new observations, the
complete conditional fitted distribution must be taken into account
rather than just the conditional mean.

The elegance of the COM-Poisson regression model lies in its ability to
address applications containing a wide range of dispersion in a
parsimonious way. While the negative binomial model is a popular
resource for count data applications where over-dispersion exists, it
cannot address problems where data are under-dispersed. The RGPR
formulation offers more flexibility in its ability to handle data
dispersion, yet it is limited in the level of under-dispersion that it
can capture. We have shown that, in such cases, the COM-Poisson
regression does not encounter such difficulties and produces reasonable
fitted models. The COM-Poisson regression has the flexibility even in
the extreme case of a binary response, where it reduces to a logistic
regression in theory, and produces identical estimates and predictors
in practice.

Our regression model is similar to the Bayesian formulation used by
\citeauthor{borle} (\citeyear{borle,borle2007}), \citet{borle2006},\break
\citet{boatwright}, \ \ \citet{kalyanam}
and that by \citet{Lord2008} in terms of the generated estimated parameters. It differs
from the Bayesian\break formulation, however, both conceptually [in terms of
the link function of \citet{Lord2008} and the estimation method] and
practically (with regard to run time). Although the Bayesian
implementation allows for the incorporation of prior information in the
form of prior parameter distributions [e.g., see \citet{kadane2005}],
the benefit of such information is useful only when informative priors
are used and when the sample size is small. Second, specifying
meaningful priors on the $\beta$ coefficients is not straightforward,
as it requires an understanding of the function ${\bolds{\lambda}
}^{1/ \nu}$, which is not equal to the mean. Software implementation
also differentiates these models because our formulation relies on
traditional estimation methods for exponential family distributions:
estimation, inference, and diagnostics can be programmed in most
statistical software packages in a straightforward manner. From a
computational point of view, although the $Z$ function requires
approximation (because it is an infinite sum), in practice, a simple
truncation of the sum performs well.

A potential restricting factor in our current COM-Poisson regression
formulation is that it assumes a constant dispersion level across all
observations. This is similar to the classic homoscedasticity
assumption in linear regression. A possible enhancement is to allow $\nu
$ to be observation-dependent (and to model it as a function of
covariates as well). In our COM-Poisson regression formulation such an
extension still maintains the structure of an exponential family,
unlike that of the generalized Poisson regression of \citet{famoye}, for example.

The relationship between the associated fitted mean bands and\break the
estimated data dispersion is nicely illustrated in accordance with\break \citet{mccullagh}. Further work is needed to investigate their impact on Type
I errors associated with hypothesis testing about the slope, or slope
coverage. In addition, this work introduces several questions regarding
sample size, which, although easily overcome by using bootstrap,
present interesting research questions.

Finally, while not presented in this work, simulations were performed
to demonstrate the accuracy of the estimation process, as well as that
of the hypothesis testing procedure. R code for simulating
COM-Poisson data is also available at
\href{http://www9.georgetown.edu/faculty/kfs7/research}{www9.georgetown.edu/faculty/kfs7/research}.

\section*{Acknowledgments}
The authors thank Seth Guikema and Dominique Lord for supplying the
Toronto crash data to allow for method comparison. The authors also
thank Jay Kadane and the reviewers for their helpful and insightful comments.

%Count Data''}
%[url]{http://www9.georgetown.edu/faculty/kfs7/research/AOAS306supplement.pdf}
%

\begin{supplement}[id=suppA]
\sname{Supplementary Materials}
\slink[doi]{10.1214/09-AOAS306SUPP}
\slink[url]{http://lib.stat.cmu.edu/aoas/306/supplement.pdf}
\sdescription{Materials include details of the iterative reweighted
least squares estimation, the Fisher information matrix components
associated with the COM-Poisson coefficients, the full airfreight data
set and diagnostics under various regression models for the airfreight
and crash data, and additional logistic regression examples.}
\end{supplement}

\printaddresses

\end{document}